\documentclass{article}
\usepackage{arxiv}
\usepackage{graphicx,url}
\usepackage[utf8]{inputenc}
\usepackage[english]{babel}
\usepackage[acronym,nonumberlist,nogroupskip,noredefwarn]{glossaries}

\usepackage[utf8]{inputenc} 
\usepackage[T1]{fontenc}    
\usepackage{booktabs}       
\usepackage{amsfonts}       
\usepackage{nicefrac}       
\usepackage{microtype}      
\usepackage{lipsum}		

\usepackage{doi}

\usepackage{graphicx}

\usepackage{listings}
\usepackage{cite}
\usepackage{amsmath,amssymb,amsfonts}
\usepackage{algorithmic}
\usepackage{graphicx}
\usepackage{textcomp}
\usepackage{xcolor}
\usepackage{placeins}
\usepackage{breqn}
\usepackage{makecell}
\usepackage{booktabs}
\usepackage{listings}
\usepackage{subcaption}
\usepackage{url}
\usepackage{hyperref}
\usepackage[multiple]{footmisc}
\usepackage{flushend}
\usepackage{adjustbox}  
\usepackage[inline]{enumitem}

\renewcommand{\shorttitle}{}

\hypersetup{
pdftitle={DATA ANALYSIS OF CLOUD VIRTUALIZATION EXPERIMENTS},
pdfsubject={cs.NI},
pdfauthor={Pedro Carmo}
}

\title{Data analysis of cloud virtualization experiments}

\author{%
    \textbf{Pedro R. X. do Carmo} \\
    Centro de Informática (CIn) \\
    Grupo de Pesquisa em Redes e Telecomunicações\\(GPRT)\\
    Universidade Federal de Pernambuco (UFPE) \\
    Recife, Brasil \\
    \texttt{pedro.carmo@gprt.ufpe.br}
    \and
    \textbf{Eduardo Freitas} \\
    Centro de Informática (CIn) \\
    Grupo de Pesquisa em Redes e Telecomunicações\\(GPRT)\\
    Universidade Federal de Pernambuco (UFPE) \\
    Recife, Brasil \\
    \texttt{eduardo.freitas@gprt.ufpe.br}
    \and
    \\
    \textbf{Assis T. de Oliveira Filho} \\
    Centro de Informática (CIn) \\
    Grupo de Pesquisa em Redes e Telecomunicações\\(GPRT)\\
    Universidade Federal de Pernambuco (UFPE) \\
    Recife, Brasil \\
    \texttt{assis.tiago@gprt.ufpe.br}
    \and
     \\
    \textbf{Judith Kelner} \\
    Centro de Informática (CIn) \\
    Grupo de Pesquisa em Redes e Telecomunicações\\(GPRT)\\
    Universidade Federal de Pernambuco (UFPE) \\
    Recife, Brasil \\
    \texttt{jk@gprt.ufpe.br}
    \and
    \\
    \textbf{Djamel Sadok} \\
    Centro de Informática (CIn) \\
    Grupo de Pesquisa em Redes e Telecomunicações\\(GPRT)\\
    Universidade Federal de Pernambuco (UFPE) \\
    Recife, Brasil \\
    \texttt{jamel@cin.ufpe.br}
}

\begin{document}
\maketitle
\begin{abstract}
The cloud computing paradigm underlines data center and telecommunication infrastructure design. Heavily leveraging virtualization, it slices hardware and software resources into smaller software units for greater flexibility of manipulation. Given the considerable benefits, several virtualization forms, with varying processing and communication overheads, emerged, including \textit{Full Virtualization} and \textit{OS Virtualization}. As a result, predicting packet throughput at the data plane turns out to be more challenging due to the additional virtualization overhead located at CPU, I/O, and network resources. This research presents a dataset of active network measurements data collected while varying various network parameters, including CPU affinity, frequency of echo packet injection, type of virtual network driver, use of CPU, I/O, or network load, and the number of concurrent VMs. The virtualization technologies used in the study include KVM, LXC, and Docker. The work examines their impact on a key network metric, namely, end-to-end latency. Also, it builds data models to evaluate the impact of a cloud computing environment on packet round-trip time. To explore data visualization, the dataset was submitted to pre-processing, correlation analysis, dimensionality reduction, and clustering. In addition, this paper provides a brief analysis of the dataset, demonstrating its use in developing machine learning-based systems for administrator decision-making.
\end{abstract}

\section{Introduction}
\label{sec:intro}

In recent years, the widespread use of cloud computing embraced a trend for using service-oriented architecture (SOA). It allows on-demand elastic allocation and sharing of hardware and software resources. Cloud computing offers a reusable and distributed computing infrastructure, reducing investment and infrastructure costs \cite{BELLO2021103441}. In summary, Cloud Computing provides access to configurable computing resources that can be made available quickly, easily, and, more importantly, on demand.

Virtualization is the driving engine that powers Cloud Computing data centers, networks, and telecommunication systems infrastructure. Network devices such as firewalls, routers, and switches have their software counterparts implemented using Virtual Network Functions (VNFs). These are software-based and designed to execute over off-the-shelf hardware. For instance, unlike previous generations of mobile architectures, 5G follows a Service-Based Architecture (SBA) comprised of VNFs. In other words, virtualization has become ubiquitous to new processing and communication architectures. Despite the many benefits, virtualization imposes additional processing overhead that affects network performance. The additional virtualization layers, including the virtual machines themselves, virtual network drivers, type of interaction with the host operating system, and disk access, are all factors that affect the time it takes for packet processing and transmission. As a result, one may observe changes to a sometimes unstable network behavior that can be traced back to the impact of using virtualization. In this paper, we shed some light on the factors behind such poor performance and analyze their impact on end-to-end packet latency.

Cloud systems are known to host a wide range of services, such as databases, web services, and network functions. Cloud applications behave differently. Some may rely on intensive CPU resources, while others mainly consume disk I/O or network resources. Since virtualization shares such resources among the hosted services, their networking performance is consequently affected given that packet processing requires CPU, disk I/O and network resources. In addition, several virtualization technologies can be used when creating virtualized applications, each using a different approach and, therefore, ultimately, a different impact on network performance, as will be shown in our experiments. As a result, Cloud administrators must consider these factors when deciding which type of virtualization technology to use. An unplanned configuration setup may lead to heavy penalties suffered by packet processing at the data plane.

In this work, we describe a dataset extracted from experiments conducted using active measurement over a network while varying several relevant parameters to emulate real Cloud Computing scenarios and their applications. A special testbed was created to assess the impact of a given virtualization technology on network performance. Data related to network performance was then collected from a testbed infrastructure running three distinct virtualization technologies: kernel-based virtual machines (KVM), Linux containers (LXC), and Docker containers. Our goal is to assess their packet processing overhead and evaluate their impact on end-to-end service latency. First, we describe the adopted testbed. Next, we present the collected dataset and analyze it using statistical and data science methods; finally, these data are used to create machine learning models designed to assist administrators in making decisions, including choosing the right virtualization technology based on the actual characteristics and requirements of the scenario.

The remainder of the paper is organized as follows: Section \ref{sec:related} brings related work relevant to our work. Section \ref{sec:testbed} presents the adopted testbed and how the data was collected. Section \ref{sec:scenario} describes our scenario, details the dataset, and introduces its processing. Section \ref{sec:analisys} describes the data analysis and shows examples of use cases of the data; more specifically, regression models built with the data are presented. Finally, Section \ref{sec:conc} concludes the article and discusses future work.
\section{Related Works}
\label{sec:related}
This paper is first to provide a dataset with experimental data from a testbed created with different virtualization technologies. It covers various virtualization scenarios, combines several network parameters and measures performance metrics such as end-to-end latency and CPU usage. This section reviews some works that also carry out experiments as well as approaches similar to our research.

The paper in \cite{7095802} compares performance of traditional virtual machine implementations to that of containers; more specifically, its authors examine the use of KVM and Docker technologies. The results show that Docker virtualization performs similarly or better in almost all the conducted tests. The used comparison scenario, however, is seen as limited. It explores a reduced number of configurations and their combinations. Those authors also do not provide the data used; they only provide the code in the form of a script for the reader interested in performing their experiments.

The work presented in \cite{10.1007/978-3-030-05057-3_5} applies a data science analysis approach in the context of NFV Infrastructure. The paper describes a Kubernetes-based testbed and collects Infrastructure as a Service (IaaS) anomaly data to build an anomaly database for NFV applications. The authors collect data from three categories of anomalies and label them. They also apply machine learning algorithms to find outliers. They built the dataset and made it publicly available to the research community. Our research adopts a similar approach to theirs. But unlike it, we focus on the types of virtualization and how they perform under specific scenarios, analyzing performance through Round-Trip time (RTT) and CPU usage metrics.

Finally, in a previous work \cite{FILHO202273}, we performed an experimental evaluation using KVM, LXC, and Docker using the dataset presented in this work. There, we compare the virtualization types by combining several parameters, such as CPU affinity, the frequency of injection of measurement echo packets, the type of virtual network driver, CPU, I/O, or network overload, and the number of background VMs. By analyzing the obtained data, we found that virtualization interferes with the RTT accuracy. In data centers with intensive I/O characteristics, KVM VMs using a VirtIO network driver achieve good results, with RTT overhead close to zero. In a high network load data center, Docker technology outperforms the others exhibiting minor variation compared to other types of virtualization. The work highlights that CPU Affinity does not significantly improve network performance. 
Furthermore, network load when applied to the receiver server generates worse results in terms of RTT. KVM achieves its best results when combined with the VirtIO driver, also in terms of end-to-end latency. Lastly, Docker virtualization outperforms LXC in almost all the adopted scenarios. In the referenced article, raw data analysis was performed through empirical analysis and the generation of graphs, while statistical methods focused on scenario analysis. Despite the many results, the present paper further expands this previous experimental analysis by performing a deeper evaluation using the same data set. Our primary focus is to draw conclusions based on data science concepts using statistical analysis and machine learning, making the dataset available and discussing the data, generating insights and findings that other researchers can use in their respective usage scenarios. Nonetheless, we also emphasize that the same test bench and data are used in both works.

\section{Testbed and Data Collection}
\label{sec:testbed}

In this section, we describe the testbed utilized to conduct the planned experiments. It emulates a virtual Cloud Computing environment and consists of two server hosts connected through a P2P topology. We distinguish among two types of virtual machines: main and background. Each one of the two servers hosts: a) main virtual machines (VMs) that measure network performance and; b) background VMs that generate resource overhead. The latter VMs emulate a Cloud Computing environment where multiple VMs, compete for resources, run concurrently and perform intense resource utilization.

\begin{figure}[ht!]
\centering
\includegraphics[width=80mm]{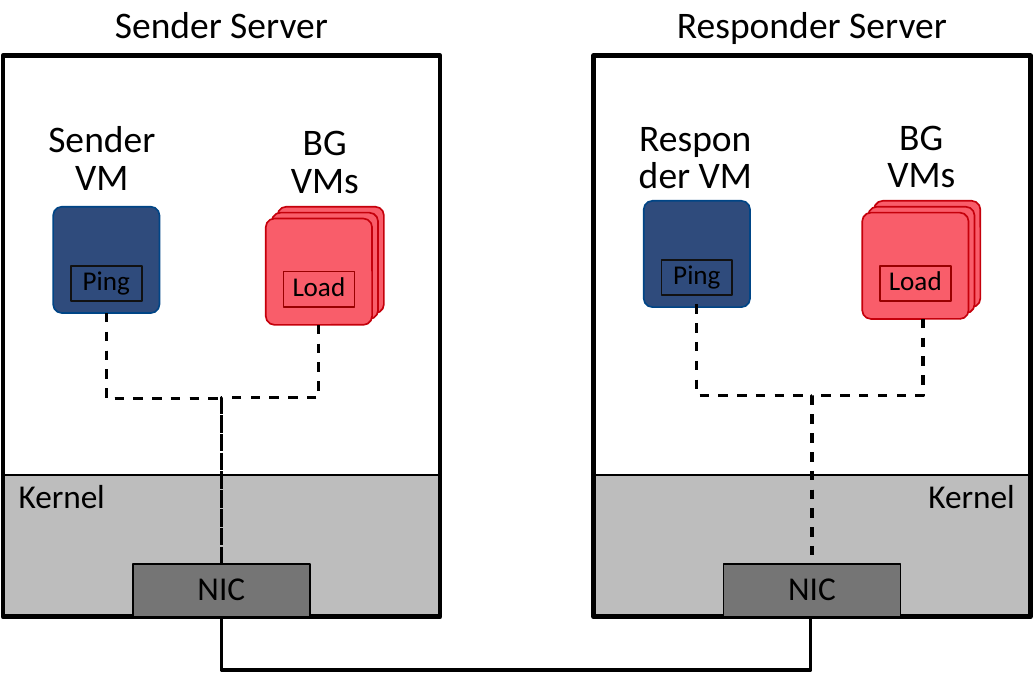}
\caption{Testbed. Adapted from \cite{FILHO202273}.} 
\label{fig:testbed}
\end{figure}

The experiments were initiated by starting the ``main VMs'', followed by the ``background VMs''. The evaluation metrics, end-to-end latency in the form of Round-Trip Time (RTT), and CPU usage were then captured. RTT was measured using the ping tool on the ``main VMs'', where the ``Sender Host'' sent the ping packets, and the ``Responder Host'' received and responded to them. RTT was calculated when the ping packet returned to the Sender VM. For statistical significance, a total of 1000 RTT samples were obtained for each experiment. CPU usage was measured using the \textit{mpstat} tool, executing on both servers. Figure~\ref{fig:testbed} illustrates a visual representation of the testbed.

The experiments were divided into two parts. Initial experiments used the KVM hypervisor whereas the second set of experiments used container-based VMs, namely, LXC and Docker. In addition, various system configurations and scenario factors were combined and varied to construct a comprehensive and representative dataset. The number of background VMs varied from 0 to 128, with 0 representing the baseline scenario without resource competition. This workload VMs compete for disk I/O, CPU, or network resources; such load location could be applied either at the ''Sender Server'' or the ''Responder Server''. The echo packet sending rate, used for the active measurement of latency, followed a frequency of 1, 1,000, or 1,000,000 packets per second (or 1, 1kHz and 1Mhz). Further details about the parameters used in the experiments and their respective value ranges can be found in Table~\ref{Tab:param}.

At the end of the experiments, several text (.txt) files were generated, containing RTT information for all sent ping packets and CPU usage information for each experiment. This data sets the first step for network administrators to analyze, draw meaningful insights, and make informed decisions regarding the best configuration for their virtualization environment and selecting the most suitable one for their operations. The following section describes the actual data treatment and dataset construction.

\section{Scenario and Dataset} 
\label{sec:scenario}
This study gathered data from the Linux operating system using the open-source tools, tcpdump and mpstat. The raw data collected was saved in text files with a ``.txt" extension. A Python script was employed to extract the data from these files and create a CSV table, which was used for pre-processing and creating predictive models and groupings. The dataset is publicly available and can be accessed in \cite{6459234}. An illustration of the dataset can be seen in Table~\ref{Tab:dataset}, which provides a few sample rows.

\begin{table*}[htb]
\resizebox{\textwidth}{!}{\begin{tabular}{|c|c|c|c|c|c|c|c|c|c|c|}
\hline
\textbf{row\_id} & \textbf{Vm\_Number} & \textbf{CPUaffinity} & \textbf{LoadLocation} & \textbf{LoadType} & \textbf{Frequency} & \textbf{Driver} & \textbf{VirtType} & \textbf{CPU\_HS\_usage} & \textbf{CPU\_HR\_usage} & \textbf{RTT\_Ping} 
\\ \hline
0 & 4 & on & HS & NET & 1000000.0 & RTL8139 & kvm  & 2.30 & 2.33 & 1.300    
\\ \hline
0 & 4 & on & HS & NET & 1000000.0 & RTL8139 & kvm  & 2.30 & 2.33 & 0.862    
\\ \hline
... & ... & ... & ... & ... & ... & ... & ... & ... & ... & ... 
\\ \hline
1609463 & 128 & off & HR & IO & 1000.0 & 0 & docker & 9.35 & 99.50 & 0.575 
\\ \hline
\end{tabular}}
\caption{Dataset}
\label{Tab:dataset}
\end{table*}

\subsection{Dataset}
\begin{figure}[ht!]
\centering
\includegraphics[width=50mm]{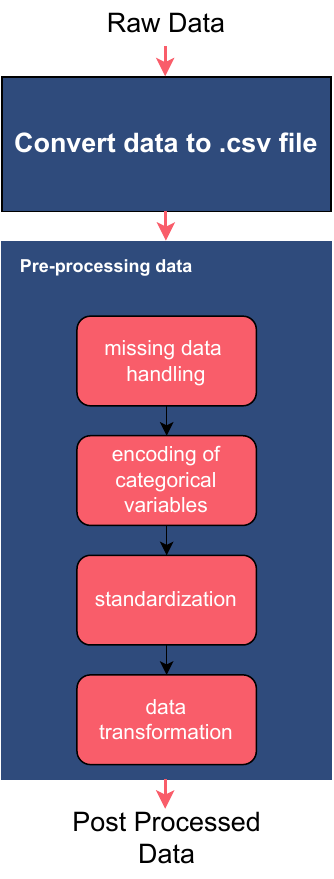}
\caption{Data processing steps} 
\label{fig:processing}
\end{figure}

The experiments conducted over the testbed produced a dataset with approximately 1,609,463 (or more than a million) entries. The dataset includes seven columns that represent the characteristics of a given network scenario. One of these columns measures the RTT performance metric, represented by the variable ``RTT\_ping". Two additional columns display the CPU usage, in percentage values, at both the sender and responder hosts during the experiments. The dataset also includes parameters encoded as integers, floats, boolean, or strings. Table~\ref{Tab:param} outlines each parameter's type, description, and possible values. A pre-processing step is necessary before using this data as input for machine learning algorithms. Figure~\ref{fig:processing} illustrates the adopted data processing flowchart.

\begin{table*}[htb]
\resizebox{\textwidth}{!}{\begin{tabular}{|c|c|c|c|}
\hline
\textbf{Parameter}      & \textbf{Description}                                                                                                                                                                           & \textbf{Type of Data} & \textbf{Values Ranges}                                                                  \\ \hline
\textbf{Vm\_number}     & Number of Background Virtual Machines                                                                                                                                                          & Integer               & \begin{tabular}[c]{@{}c@{}}0,1,2,4,8,16,\\ 32,64,128\end{tabular}                       \\ \hline
\textbf{CPUaffinity}   & \begin{tabular}[c]{@{}c@{}}Identifies whether a CPU has been \\ assigned to process packets\end{tabular}                                                                                       & Boolean               & on/off                                                                              \\ \hline

\textbf{LoadLoaction}  & \begin{tabular}[c]{@{}c@{}}The impact of workload that each background VM \\ performs also differs according to whether it is running \\ at the Sender Server (HS) or Responder Server (HR).\end{tabular} & String                & \begin{tabular}[c]{@{}c@{}}``HS" or\\ ``HR"\end{tabular}         \\ \hline
\textbf{LoadType}      & \begin{tabular}[c]{@{}c@{}}The workload each background VM performs \\ targets a different hardware resource.\\ These include either CPU, network or disk I/O load\end{tabular}                & String                & \begin{tabular}[c]{@{}c@{}}``CPU", ``IO" or\\  ``NET"\end{tabular}                      \\ \hline
\textbf{Frequency}      & The frequency of sending packets in pps (packets per second)                                                                                                                                                             & Float               & 1.0, 1000.0, 1000000.0                                                                       \\ \hline
\textbf{Driver}          & Virtual Network Driver used with KVM                                                                                                                                                           & String                & \begin{tabular}[c]{@{}c@{}}``RTL8139", ``E1000",\\ ``VirtIO", ``VMXNET3" and\\``0"(if VirtType is not KVM")\end{tabular} \\ \hline
\textbf{VirtType}       & Type of virtualization used                                                                                                                                                                    & String                & ``LXC", ``KVM", ``DOCKER"                                                               \\ \hline
\textbf{CPU\_HS\_usage} & CPU usage on sender host in percentage                                                                                                                                                         & Float                 & 0.0 to 100.0                                                                            \\ \hline
\textbf{CPU\_HR\_usage} & CPU usage on responder host in percentage                                                                                                                                                      & Float                 & 0.0 to 100.0                                                                            \\ \hline
\textbf{RTT\_Ping}      & Round-Trip Time                                                                                                                                                                                & Float                 & any non-negative float number                                                                         \\ \hline
\end{tabular}}
\caption{Dataset parameters}
\label{Tab:param}
\end{table*}

\subsection{Pre-processing} \label{subsec:preprocess}

The pre-processing step is essential for ensuring the accuracy and quality of the results in our analysis. Incorrect data can lead to distorted results, as it makes pattern detection difficult and leads to poorly performing models \cite{jain2020overview, geron2019hands}. This step involves transforming and encoding variables, extracting features, and removing useless or incomplete data.

First, missing data needs to be handled. The \emph{``Driver"} parameter column only contains non-zero values when the \texttt{``VirtType"} column is equal to \texttt{``KVM"}. To address this issue, we combined the \emph{``Driver"} and \texttt{``VirtType"} columns to create a new feature, resulting in possible values of \texttt{``KVM\_virtio," ``KVM\_rtl8139," ``KVM\_e1000," ``KVM\_vmxnet3," ``LXC," and ``DOCKER"}.

Next, the encoding is performed using One-Hot Encoding \cite{cohen2014applied}, a technique that transforms categorical data into a binary representation. This method creates a column for each possible value of the categorical variable and adds a 1 in the appropriate column. The categorical columns in our dataset, \texttt{LoadLocation, LoadType, Driver, and VirtType}, were all encoded using One-Hot Encoding. The new columns resulting from this step are \texttt{CPU, IO, NET, HS, HR, DOCKER, LXC, KVM\_E1000, KVM\_VIRTIO, KVM\_RTL8139, KVM\_VMXNET3}, all with possible values of 1 or 0, where 1 indicates the presence of the parameter and 0 indicates its absence.

In addition, the non-categorical columns were normalized using the z-score standardization method, transforming the data into a distribution with a mean of 0 and a standard deviation of 1. This method was used to avoid any adverse effects on the performance of some algorithms and to prevent the problem of combining values with different scales, such as the ``Frequency" column ranging from 1 to $10^6$ and the \emph{VM\_number} column ranging from 0 to 128. The values are calculated using the equation \ref{eq:zscore}, shown below:

\begin{equation}
z = \frac{x - \mu}{\sigma}
\label{eq:zscore}
\end{equation}

where $x$ is the original value, $\mu$ is the mean of the data, and $\sigma$ is the standard deviation.

Once the data transformation step is carried out, where the types of variables are adapted to facilitate analysis and experiments, all One-hot encoding columns are set to Integer values. In column \texttt{CPU\_Affinity}, the value  ``on" is transformed to 1 and the value ``off" to 0, both integers. The \texttt{RTT\_ping, CPU\_HR\_usage, CPU\_HS\_usage, Frequency} columns remain encoded as Float whereas the \texttt{VM\_number} column remains Integer based.

\begin{table}[tbp]
\centering
\begin{tabular}{|c|c|c|c|c|}
\hline
... & ... & \textbf{CPU} & \textbf{IO} & \textbf{NET} \\ \hline
... & ... & 1            & 0           & 0            \\ \hline
... & ... & ...          & ...         & ...          \\ \hline
... & ... & 0            & 1           & 0            \\ \hline
... & ... & 0            & 0           & 1            \\ \hline
\end{tabular}
\caption{One-hot-enconding example}
\label{Tab:onehot}
\end{table}
\section{Data Analysis and Use case} \label{sec:analisys}
In this Section, we analyze the dataset and discuss its potential applications.

\subsection{Statistical Analysis}
\begin{figure}[ht!]
\centering
\includegraphics[width=80mm]{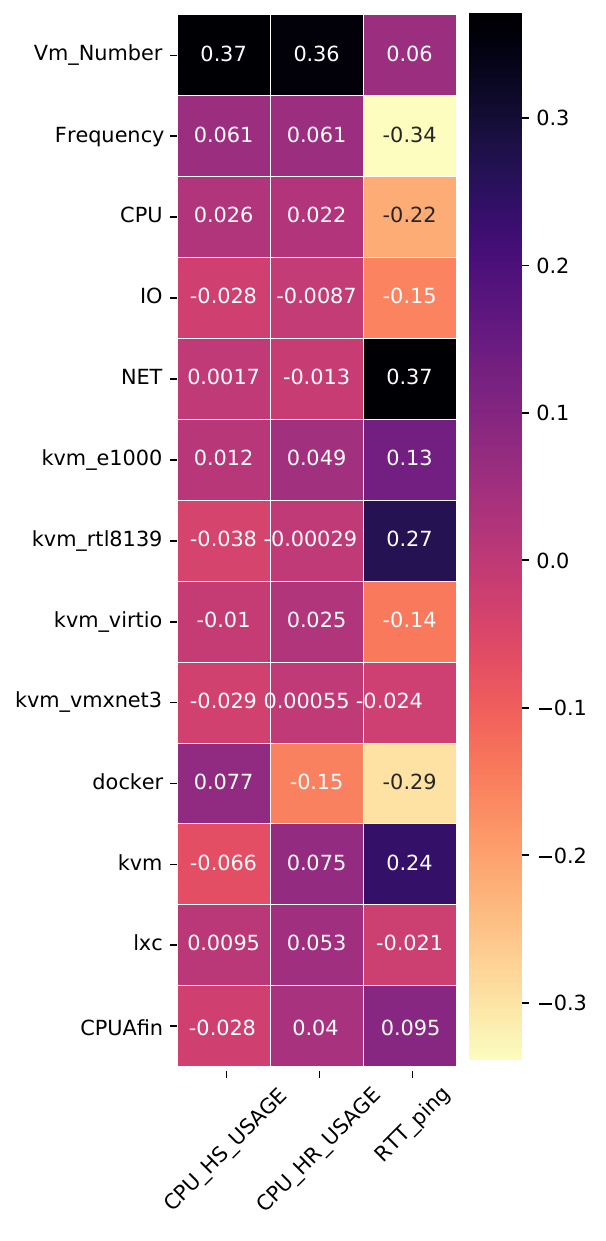}
\caption{Correlation} 
\label{corr}
\end{figure}

The data analysis performed in this study is based on the output of the previous pre-processing stage. A preliminary analysis uses Spearman's rank-order correlation, a non-parametric version of Pearson correlation, as described in \cite{zar2014spearman}. The correlation measures the strength and direction of the association between two ranked variables, with values ranging from -1 to 1. A negative value indicates an inverse correlation, a positive value indicates a direct correlation, and a value of 0 indicates a lack of correlation. The observed correlation between the scenario variables, which define the scenario, and the performance variables, to evaluate the scenario performance, is displayed in Figure~\ref{corr}.

\paragraph{\textbf{Discussion}}
\medskip The results of the initial test lead to several conclusions. As seen in Figure~\ref{corr}, the number of VMs running in the background significantly impacts the CPU usage of both the sender and responder hosts. On the other hand, the other variables have little effect on CPU consumption. Regarding CPU usage, using Docker leads to decreased CPU usage on the responder host. Furthermore, it is observed that Stress NET (more VMs to generate network background traffic??) has a strong relationship with increased RTT. Using the VirtIO driver significantly enhances the performance of KVM virtualization, unlike the rtl8139 driver, which reduces its efficiency. This is because VirtIO drivers provide direct (paravirtualized) access to devices and peripherals for virtual machines (VMs), making them faster than emulated drivers, which are slower. Additionally, Docker virtualization technology has the best performance for both CPU usage and latency metrics. Interestingly, our analysis reveals that RTT decreases as the measurement frequency using ping echo packets increases. This effect is attributed to shorter CPU wake-up times and the system's preference for interrupt-based delivery, which is more efficient for packets at higher frequencies \cite{FILHO202273}. Additionally, the number of VMs in the background does not impact the RTT metric, and using CPU Affinity does not improve either metric.

\subsection{Clustering and PCA}

\begin{figure}[ht!]
\centering
\includegraphics[width=80mm]{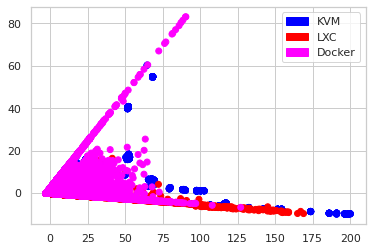}
\caption{Principal Component Analysis Plot. The X-axis represents the first principal component, and the Y-axis represents the second principal component. The data points represent the samples being analyzed. The vectors indicate the direction of maximum variance. Each color represents an analyzed technology: Docker, KVM, and LXC.} 
\label{PCA}
\end{figure}

This section introduces a strategy used for further analysis of the observed measurements. Principal Component Analysis (PCA) is a popular dimensionality reduction technique widely used in the literature \cite{abdi2010principal}. It allows us to project a high-dimensional dataset into a lower-dimensional space, thereby reducing its dimensionality and making it easier to visualize. In this study, we applied PCA to our dataset and reduced it to two dimensions, as shown in Figure~\ref{PCA}. The data points are labeled according to the three different virtualization techniques we tested: KVM, LXC, and Docker. Although there is no clear separation between the three techniques, it is evident that KVM and LXC tend to occupy a space that is distinct from the space occupied by Docker.

As a continuation of the data analysis, we next apply a clustering technique to visualize the main components of the earlier described Principal Component Analysis (PCA). Clustering is a machine learning technique that falls under unsupervised learning methods. Its objective is to divide the data into groups based on similarity. In this study, we use the KMeans algorithm \cite{electronics9081295} to separate our data into three groups, corresponding to the three virtualization groups: Docker, LXC, and KVM.

The KMeans algorithm divides the data into $K$ clusters based on their distances from $K$ centroids, representing each cluster's center. The initial position of the centroids is determined randomly, and then each data point is assigned to the cluster with the closest centroid. Finally, the position of the centroids is recalculated by taking the average of the data points in each cluster. In our case, we set $K=3$.

\begin{figure}[ht!]
\centering
\includegraphics[width=80mm]{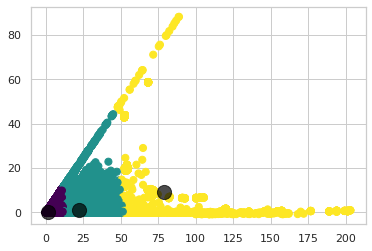}
\caption{K-Means Clustering Plot. The X-axis and Y-axis show the values of two principal components, and each data point is positioned according to its values on these two components. Data points are grouped into 3 clusters represented by different colors. The centroids of the 3 clusters are represented by a wide dot at each cluster's center.} 
\label{cluster}
\end{figure}

The result of the clustering algorithm is shown in Figure~\ref{cluster}. The wide dots in the image represent the centroids of the data. It is evident that the three groups, Docker, LXC, and KVM, can be identified. This leads us to conclude that the combined use of PCA and KMeans effectively visualizes our high-dimensional data.

\begin{table}[htbp]
\centering
\begin{tabular}{|c|c|c|}
\hline
& \textbf{Input Variables}  & \textbf{Output Variables} \\ \hline
\textbf{CPU Model} & \begin{tabular}[c]{@{}c@{}}Frequency,  Vm\_number, \\CPUaffinity, CPU, IO, \\NET, HS, HR,\\  DOCKER, LXC, KVM\_E1000, \\ KVM\_VIRTIO, KVM\_RTL8139, \\ KVM\_VMXNET3\end{tabular} & CPU\_HR\_USAGE            \\ \hline
\textbf{RTT Model} & \begin{tabular}[c]{@{}c@{}}Frequency,  Vm\_number, \\CPUaffinity, CPU, IO, \\NET, HS, HR,\\  DOCKER, LXC, KVM\_E1000, \\ KVM\_VIRTIO, KVM\_RTL8139, \\ KVM\_VMXNET3\end{tabular} & RTT\_Ping                 \\ \hline
\end{tabular}
\caption{Models}
\label{Tab:models}
\end{table}

\subsection{Application of machine learning-based regression methods}

For the scenario described in this work, machine learning techniques can provide valuable insights for network administrators in making architectural decisions, such as choosing the most suitable virtualization technology or estimating the expected average expected latency for a specific scenario.

We present two examples of models constructed using this dataset to estimate the CPU usage of the Responder Host (CPU Model) and the average RTT (RTT Model) based on the scenario's variables. These models can help predict the behavior of scenarios not present in the dataset, for instance, scenarios with a frequency range between 1Kpps and 1Mpps or those with a different number of virtual machines compared to those listed in Table~\ref{Tab:param}. By utilizing these models, network administrators can make informed decisions on the virtualization technology that best aligns with their design goals. Regression techniques are used to build these models, with the help of the PyCaret library~\cite{PyCaret}. The variables used in constructing the models are shown in Table~\ref{Tab:models}.

In building these models, we use post-processed data, which has undergone normalization through one-hot encoding. This step is crucial for the model to learn the patterns in the data effectively.

\subsubsection{\textbf{Use case methodology}}
A methodology was established to construct two models from the dataset. Firstly, the dataset was divided into 80\% training, 10\% testing, and 10\% validation. Using the PyCaret library, up to 19 regression techniques were applied to the training set to create models. The best technique was selected based on several evaluation metrics, including Mean Absolute Error (MAE), Mean Squared Error (MSE), Root Mean Squared Error (RMSE), Coefficient of Determination (R2), Root Mean Squared Log Error (RMSLE), Mean Absolute Percentage Error (MAPE), and Training Time (TT). The chosen technique was then retrained, focusing next on hyperparameter tuning, using 5-fold cross-validation.

\paragraph{\textbf{RTT Model results}}

The RTT Model was constructed to estimate the average RTT for a given scenario based on the input variables described in Table~\ref{Tab:models}. In the first stage of training, 17 regression models were used, and the results are presented in Table~\ref{Tab:rtt_model}. From these results, it can be seen that models based on decision trees outperformed the others. Therefore, the model based on the Random Forest technique was selected for the next stage of training due to its superior performance in terms of the three error metrics: Mean Absolute Error (MAE), Mean Squared Error (MSE), and Root Mean Squared Error (RMSE).

\begin{table*}[ht]
\centering
\resizebox{\textwidth}{!}{\begin{tabular}{|llllllll|}
\hline
\multicolumn{8}{|c|}{\textbf{RTT MODEL}} \\ \hline
\multicolumn{1}{|l|}{\textbf{Model}} & \multicolumn{1}{l|}{\textbf{MAE}} & \multicolumn{1}{l|}{\textbf{MSE}} & \multicolumn{1}{l|}{\textbf{RMSE}} & \multicolumn{1}{l|}{\textbf{R2}} & \multicolumn{1}{l|}{\textbf{RMSLE}} & \multicolumn{1}{l|}{\textbf{MAPE}} & \textbf{TT (Sec)} \\ \hline
\multicolumn{1}{|l|}{\textbf{Decision Tree Regressor}} & \multicolumn{1}{l|}{\textbf{1.1343}} & \multicolumn{1}{l|}{\textbf{14.8711}} & \multicolumn{1}{l|}{\textbf{3.8560}} & \multicolumn{1}{l|}{\textbf{0.7037}} & \multicolumn{1}{l|}{\textbf{0.2978}} & \multicolumn{1}{l|}{\textbf{0.1771}} & \textbf{3.3290} \\ \hline
\multicolumn{1}{|l|}{\textbf{Extra Trees Regressor}} & \multicolumn{1}{l|}{\textbf{1.1343}} & \multicolumn{1}{l|}{\textbf{14.8711}} & \multicolumn{1}{l|}{\textbf{3.8560}} & \multicolumn{1}{l|}{\textbf{0.7037}} & \multicolumn{1}{l|}{\textbf{0.2978}} & \multicolumn{1}{l|}{\textbf{0.1771}} & 128.8010 \\ \hline
\multicolumn{1}{|l|}{\textbf{Random Forest Regressor}} & \multicolumn{1}{l|}{\textbf{1.1342}} & \multicolumn{1}{l|}{\textbf{14.8699}} & \multicolumn{1}{l|}{\textbf{3.8558}} & \multicolumn{1}{l|}{\textbf{0.7037}} & \multicolumn{1}{l|}{\textbf{0.2978}} & \multicolumn{1}{l|}{\textbf{0.1771}} & 132.3980 \\ \hline
\multicolumn{1}{|l|}{\textbf{CatBoost Regressor}} & \multicolumn{1}{l|}{1.1454} & \multicolumn{1}{l|}{15.2269} & \multicolumn{1}{l|}{3.9018} & \multicolumn{1}{l|}{0.6966} & \multicolumn{1}{l|}{0.2989} & \multicolumn{1}{l|}{0.1780} & 164.1080 \\ \hline
\multicolumn{1}{|l|}{\textbf{Extreme Gradient Boosting}} & \multicolumn{1}{l|}{1.1490} & \multicolumn{1}{l|}{15.2950} & \multicolumn{1}{l|}{3.9106} & \multicolumn{1}{l|}{0.6952} & \multicolumn{1}{l|}{0.2989} & \multicolumn{1}{l|}{0.1781} & 123.2300 \\ \hline
\multicolumn{1}{|l|}{\textbf{Light Gradient Boosting Machine}} & \multicolumn{1}{l|}{1.1831} & \multicolumn{1}{l|}{17.1895} & \multicolumn{1}{l|}{4.1445} & \multicolumn{1}{l|}{0.6575} & \multicolumn{1}{l|}{0.3038} & \multicolumn{1}{l|}{0.1791} & 10.0270 \\ \hline
\multicolumn{1}{|l|}{\textbf{AdaBoost Regressor}} & \multicolumn{1}{l|}{1.8861} & \multicolumn{1}{l|}{42.1433} & \multicolumn{1}{l|}{6.4909} & \multicolumn{1}{l|}{0.1606} & \multicolumn{1}{l|}{0.5289} & \multicolumn{1}{l|}{0.4500} & 40.0160 \\ \hline
\multicolumn{1}{|l|}{\textbf{Linear Regression}} & \multicolumn{1}{l|}{2.2643} & \multicolumn{1}{l|}{50.1294} & \multicolumn{1}{l|}{7.0793} & \multicolumn{1}{l|}{0.0015} & \multicolumn{1}{l|}{0.7203} & \multicolumn{1}{l|}{0.6261} & 6.5000 \\ \hline
\multicolumn{1}{|l|}{\textbf{Bayesian Ridge}} & \multicolumn{1}{l|}{2.2646} & \multicolumn{1}{l|}{50.1469} & \multicolumn{1}{l|}{7.0806} & \multicolumn{1}{l|}{0.0011} & \multicolumn{1}{l|}{0.7206} & \multicolumn{1}{l|}{0.6253} & 2.1730 \\ \hline
\multicolumn{1}{|l|}{\textbf{Ridge Regression}} & \multicolumn{1}{l|}{2.2646} & \multicolumn{1}{l|}{50.1467} & \multicolumn{1}{l|}{7.0806} & \multicolumn{1}{l|}{0.0011} & \multicolumn{1}{l|}{0.7206} & \multicolumn{1}{l|}{0.6253} & 1.2770 \\ \hline
\multicolumn{1}{|l|}{\textbf{Least Angle Regression}} & \multicolumn{1}{l|}{2.2646} & \multicolumn{1}{l|}{50.1468} & \multicolumn{1}{l|}{7.0806} & \multicolumn{1}{l|}{0.0011} & \multicolumn{1}{l|}{0.7206} & \multicolumn{1}{l|}{0.6253} & 1.4100 \\ \hline
\multicolumn{1}{|l|}{\textbf{Huber Regressor}} & \multicolumn{1}{l|}{2.2510} & \multicolumn{1}{l|}{50.1597} & \multicolumn{1}{l|}{7.0815} & \multicolumn{1}{l|}{0.0008} & \multicolumn{1}{l|}{0.7288} & \multicolumn{1}{l|}{0.6099} & 9.1410 \\ \hline
\multicolumn{1}{|l|}{\textbf{Orthogonal Matching Pursuit}} & \multicolumn{1}{l|}{2.4137} & \multicolumn{1}{l|}{53.1145} & \multicolumn{1}{l|}{7.2872} & \multicolumn{1}{l|}{-0.0581} & \multicolumn{1}{l|}{0.8037} & \multicolumn{1}{l|}{0.7681} & 1.3610 \\ \hline
\multicolumn{1}{|l|}{\textbf{Lasso Regression}} & \multicolumn{1}{l|}{2.4605} & \multicolumn{1}{l|}{55.0244} & \multicolumn{1}{l|}{7.4171} & \multicolumn{1}{l|}{-0.0962} & \multicolumn{1}{l|}{0.8813} & \multicolumn{1}{l|}{0.7342} & 1.4120 \\ \hline
\multicolumn{1}{|l|}{\textbf{Elastic Net}} & \multicolumn{1}{l|}{2.4605} & \multicolumn{1}{l|}{55.0244} & \multicolumn{1}{l|}{7.4171} & \multicolumn{1}{l|}{-0.0962} & \multicolumn{1}{l|}{0.8813} & \multicolumn{1}{l|}{0.7342} & 1.4010 \\ \hline
\multicolumn{1}{|l|}{\textbf{Lasso Least Angle Regression}} & \multicolumn{1}{l|}{2.4605} & \multicolumn{1}{l|}{55.0244} & \multicolumn{1}{l|}{7.4171} & \multicolumn{1}{l|}{-0.0962} & \multicolumn{1}{l|}{0.8813} & \multicolumn{1}{l|}{0.7342} & 1.4910 \\ \hline
\multicolumn{1}{|l|}{\textbf{Dummy Regressor}} & \multicolumn{1}{l|}{2.4605} & \multicolumn{1}{l|}{55.0244} & \multicolumn{1}{l|}{7.4171} & \multicolumn{1}{l|}{-0.0962} & \multicolumn{1}{l|}{0.8813} & \multicolumn{1}{l|}{0.7342} & 0.8080 \\ \hline
\end{tabular}}
\caption{RTT Models}
\label{Tab:rtt_model}
\end{table*}
\begin{table*}[]
\centering
\resizebox{\textwidth}{!}{\begin{tabular}{|llllllll|}
\hline
\multicolumn{8}{|c|}{\textbf{CPU MODELS}} \\ \hline
\multicolumn{1}{|l|}{\textbf{Model}}                           & \multicolumn{1}{l|}{\textbf{MAE}}                            & \multicolumn{1}{l|}{\textbf{MSE}}                            & \multicolumn{1}{l|}{\textbf{RMSE}}                           & \multicolumn{1}{l|}{\textbf{R2}}                             & \multicolumn{1}{l|}{\textbf{RMSLE}}                          & \multicolumn{1}{l|}{\textbf{MAPE}}                           & \textbf{TT (Sec)}                       \\ \hline
\multicolumn{1}{|l|}{\textbf{Decision Tree Regressor}}         & \multicolumn{1}{l|}{\textbf{0.3032}} & \multicolumn{1}{l|}{\textbf{0.9062}} & \multicolumn{1}{l|}{\textbf{0.9519}} & \multicolumn{1}{l|}{\textbf{0.9992}} & \multicolumn{1}{l|}{\textbf{0.0840}} & \multicolumn{1}{l|}{\textbf{0.0391}} & \textbf{3.4690} \\ \hline
\multicolumn{1}{|l|}{\textbf{Extra Trees Regressor}}           & \multicolumn{1}{l|}{\textbf{0.3032}} & \multicolumn{1}{l|}{\textbf{0.9062}} & \multicolumn{1}{l|}{\textbf{0.9519}} & \multicolumn{1}{l|}{\textbf{0.9992}} & \multicolumn{1}{l|}{\textbf{0.0840}} & \multicolumn{1}{l|}{\textbf{0.0391}} & 107.3600                                \\ \hline
\multicolumn{1}{|l|}{\textbf{Random Forest Regressor}}         & \multicolumn{1}{l|}{\textbf{0.3032}} & \multicolumn{1}{l|}{0.9064}                                  & \multicolumn{1}{l|}{0.9520}                                  & \multicolumn{1}{l|}{\textbf{0.9992}} & \multicolumn{1}{l|}{\textbf{0.0840}} & \multicolumn{1}{l|}{\textbf{0.0391}} & 94.8720                                 \\ \hline
\multicolumn{1}{|l|}{\textbf{Light Gradient Boosting Machine}} & \multicolumn{1}{l|}{1.0115}                                  & \multicolumn{1}{l|}{4.1813}                                  & \multicolumn{1}{l|}{2.0445}                                  & \multicolumn{1}{l|}{0.9963}                                  & \multicolumn{1}{l|}{0.1142}                                  & \multicolumn{1}{l|}{0.0797}                                  & 4.2710                                  \\ \hline
\multicolumn{1}{|l|}{\textbf{CatBoost Regressor}}              & \multicolumn{1}{l|}{0.9073}                                  & \multicolumn{1}{l|}{5.7624}                                  & \multicolumn{1}{l|}{2.3985}                                  & \multicolumn{1}{l|}{0.9949}                                  & \multicolumn{1}{l|}{0.0946}                                  & \multicolumn{1}{l|}{0.0585}                                  & 106.5100                                \\ \hline
\multicolumn{1}{|l|}{\textbf{Extreme Gradient Boosting}}       & \multicolumn{1}{l|}{0.9535}                                  & \multicolumn{1}{l|}{6.1402}                                  & \multicolumn{1}{l|}{2.4767}                                  & \multicolumn{1}{l|}{0.9946}                                  & \multicolumn{1}{l|}{0.0965}                                  & \multicolumn{1}{l|}{0.0603}                                  & 59.6990                                 \\ \hline
\multicolumn{1}{|l|}{\textbf{AdaBoost Regressor}}              & \multicolumn{1}{l|}{11.1936}                                 & \multicolumn{1}{l|}{441.8921}                                & \multicolumn{1}{l|}{21.0114}                                 & \multicolumn{1}{l|}{0.6117}                                  & \multicolumn{1}{l|}{0.5464}                                  & \multicolumn{1}{l|}{0.5635}                                  & 39.3820                                 \\ \hline
\multicolumn{1}{|l|}{\textbf{Huber Regressor}}                 & \multicolumn{1}{l|}{12.8150}                                 & \multicolumn{1}{l|}{511.1350}                                & \multicolumn{1}{l|}{22.6082}                                 & \multicolumn{1}{l|}{0.5509}                                  & \multicolumn{1}{l|}{0.7587}                                  & \multicolumn{1}{l|}{0.9763}                                  & 6.6470                                  \\ \hline
\multicolumn{1}{|l|}{\textbf{Linear Regression}}               & \multicolumn{1}{l|}{14.1959}                                 & \multicolumn{1}{l|}{645.4639}                                & \multicolumn{1}{l|}{25.4059}                                 & \multicolumn{1}{l|}{0.4329}                                  & \multicolumn{1}{l|}{0.7886}                                  & \multicolumn{1}{l|}{0.8921}                                  & 2.3390                                  \\ \hline
\multicolumn{1}{|l|}{\textbf{Bayesian Ridge}}                  & \multicolumn{1}{l|}{14.1991}                                 & \multicolumn{1}{l|}{645.9363}                                & \multicolumn{1}{l|}{25.4152}                                 & \multicolumn{1}{l|}{0.4325}                                  & \multicolumn{1}{l|}{0.7887}                                  & \multicolumn{1}{l|}{0.8917}                                  & 2.5530                                  \\ \hline
\multicolumn{1}{|l|}{\textbf{Ridge Regression}}                & \multicolumn{1}{l|}{14.1991}                                 & \multicolumn{1}{l|}{645.9339}                                & \multicolumn{1}{l|}{25.4152}                                 & \multicolumn{1}{l|}{0.4325}                                  & \multicolumn{1}{l|}{0.7887}                                  & \multicolumn{1}{l|}{0.8917}                                  & 1.4190                                  \\ \hline
\multicolumn{1}{|l|}{\textbf{Least Angle Regression}}          & \multicolumn{1}{l|}{14.1998}                                 & \multicolumn{1}{l|}{645.9888}                                & \multicolumn{1}{l|}{25.4162}                                 & \multicolumn{1}{l|}{0.4324}                                  & \multicolumn{1}{l|}{0.7887}                                  & \multicolumn{1}{l|}{0.8917}                                  & 1.1970                                  \\ \hline
\multicolumn{1}{|l|}{\textbf{Gradient Boosting Regressor}}     & \multicolumn{1}{l|}{5.5745}                                  & \multicolumn{1}{l|}{680.9593}                                & \multicolumn{1}{l|}{26.0342}                                 & \multicolumn{1}{l|}{0.4017}                                  & \multicolumn{1}{l|}{0.2841}                                  & \multicolumn{1}{l|}{0.2415}                                  & 45.6180                                 \\ \hline
\multicolumn{1}{|l|}{\textbf{Orthogonal Matching Pursuit}}     & \multicolumn{1}{l|}{17.9799}                                 & \multicolumn{1}{l|}{1082.1288}                               & \multicolumn{1}{l|}{32.8956}                                 & \multicolumn{1}{l|}{0.0492}                                  & \multicolumn{1}{l|}{0.9987}                                  & \multicolumn{1}{l|}{1.2008}                                  & 1.0770                                  \\ \hline
\multicolumn{1}{|l|}{\textbf{Lasso Regression}}                & \multicolumn{1}{l|}{20.8854}                                 & \multicolumn{1}{l|}{1438.2111}                               & \multicolumn{1}{l|}{37.9236}                                 & \multicolumn{1}{l|}{-0.2636}                                 & \multicolumn{1}{l|}{1.3929}                                  & \multicolumn{1}{l|}{1.4613}                                  & 1.2280                                  \\ \hline
\multicolumn{1}{|l|}{\textbf{Elastic Net}}                     & \multicolumn{1}{l|}{20.8854}                                 & \multicolumn{1}{l|}{1438.2111}                               & \multicolumn{1}{l|}{37.9236}                                 & \multicolumn{1}{l|}{-0.2636}                                 & \multicolumn{1}{l|}{1.3929}                                  & \multicolumn{1}{l|}{1.4613}                                  & 1.2450                                  \\ \hline
\multicolumn{1}{|l|}{\textbf{Lasso Least Angle Regression}}    & \multicolumn{1}{l|}{20.8854}                                 & \multicolumn{1}{l|}{1438.2111}                               & \multicolumn{1}{l|}{37.9236}                                 & \multicolumn{1}{l|}{-0.2636}                                 & \multicolumn{1}{l|}{1.3929}                                  & \multicolumn{1}{l|}{1.4613}                                  & 1.3330                                  \\ \hline
\multicolumn{1}{|l|}{\textbf{Dummy Regressor}}                 & \multicolumn{1}{l|}{20.8854}                                 & \multicolumn{1}{l|}{1438.2111}                               & \multicolumn{1}{l|}{37.9236}                                 & \multicolumn{1}{l|}{-0.2636}                                 & \multicolumn{1}{l|}{1.3929}                                  & \multicolumn{1}{l|}{1.4613}                                  & 0.6770                                  \\ \hline
\multicolumn{1}{|l|}{\textbf{Passive Aggressive Regressor}}    & \multicolumn{1}{l|}{35.1884}                                 & \multicolumn{1}{l|}{85362.3373}                              & \multicolumn{1}{l|}{199.6070}                                & \multicolumn{1}{l|}{-73.7373}                                & \multicolumn{1}{l|}{1.0533}                                  & \multicolumn{1}{l|}{1.5621}                                  & 2.0520                                  \\ \hline
\end{tabular}}
\caption{CPU Models}
\label{Tab:cpu_model}
\end{table*}
Below are the hyperparameters selected in the RTT Model tuning step:

\begin{center}
\medskip
\fbox{\begin{minipage}{15em}
\centering
\textbf{RandomForest}(ccp\_alpha=0.0, criterion=`mse',\\
    max\_features=`auto',\\
    min\_impurity\_decrease=0.0, min\_samples\_leaf=1,\\
    min\_samples\_split=2,\\ min\_weight\_fraction\_leaf=0.0,\\
    n\_estimators=100,\\
    n\_jobs=-1, oob\_score=False,\\
    random\_state=123)%
\end{minipage}}
\medskip
\end{center}

Table~\ref{Tab:rf-rtt} displays the result of the second training step for the selected model. The reader can see a performance improvement achieved by most metrics.
\begin{table}[htb]
\centering
\begin{tabular}{|ccccccc|}
\hline
\multicolumn{7}{|c|}{\textbf{Random Forest}} \\ \hline
\multicolumn{1}{|c|}{\textbf{Fold}} & \multicolumn{1}{c|}{\textbf{MAE}} & \multicolumn{1}{c|}{\textbf{MSE}} & \multicolumn{1}{c|}{\textbf{RMSE}} & \multicolumn{1}{c|}{\textbf{R2}} & \multicolumn{1}{c|}{\textbf{RMSLE}} & \textbf{MAPE} \\ \hline
\multicolumn{1}{|c|}{\textbf{0}} & \multicolumn{1}{c|}{1.1527} & \multicolumn{1}{c|}{15.0791} & \multicolumn{1}{c|}{3.8832} & \multicolumn{1}{c|}{0.7015} & \multicolumn{1}{c|}{0.2992} & 0.1765 \\ \hline
\multicolumn{1}{|c|}{\textbf{1}} & \multicolumn{1}{c|}{1.1366} & \multicolumn{1}{c|}{15.1332} & \multicolumn{1}{c|}{3.8901} & \multicolumn{1}{c|}{0.6993} & \multicolumn{1}{c|}{0.2967} & 0.1763 \\ \hline
\multicolumn{1}{|c|}{\textbf{2}} & \multicolumn{1}{c|}{1.1459} & \multicolumn{1}{c|}{15.1237} & \multicolumn{1}{c|}{3.8889} & \multicolumn{1}{c|}{0.7098} & \multicolumn{1}{c|}{0.2980} & 0.1780 \\ \hline
\multicolumn{1}{|c|}{\textbf{3}} & \multicolumn{1}{c|}{1.1209} & \multicolumn{1}{c|}{14.6488} & \multicolumn{1}{c|}{3.8274} & \multicolumn{1}{c|}{0.7007} & \multicolumn{1}{c|}{0.2960} & 0.1749 \\ \hline
\multicolumn{1}{|c|}{\textbf{4}} & \multicolumn{1}{c|}{1.1232} & \multicolumn{1}{c|}{14.1276} & \multicolumn{1}{c|}{3.7587} & \multicolumn{1}{c|}{0.7006} & \multicolumn{1}{c|}{0.2994} & 0.1772 \\ \hline
\multicolumn{1}{|c|}{\textbf{Mean}} & \multicolumn{1}{c|}{\textbf{1.1359}} & \multicolumn{1}{c|}{\textbf{14.8225}} & \multicolumn{1}{c|}{\textbf{3.8497}} & \multicolumn{1}{c|}{\textbf{0.7024}} & \multicolumn{1}{c|}{\textbf{0,2979}} & \textbf{0.1766} \\ \hline
\multicolumn{1}{|c|}{\textbf{SD}} & \multicolumn{1}{c|}{0.0139} & \multicolumn{1}{c|}{0.4377} & \multicolumn{1}{c|}{0.0572} & \multicolumn{1}{c|}{0.0042} & \multicolumn{1}{c|}{0.0015} & 0.0012 \\ \hline
\end{tabular}
\caption{Random Forest-based RTT Model}
\label{Tab:rf-rtt}
\end{table}
With this model at hand, an administrator can estimate the RTT of a given scenario and determine which type of virtualization or driver he will use.

\paragraph{\textbf{CPU Model results}}

The second model constructed was the CPU Model. Its purpose is to estimate the percentage of CPU usage based on the variables listed in Table~\ref{Tab:models} for a given scenario. As shown in Table~\ref{Tab:cpu_model}, 19 regression models were utilized in the first stage of the training process. The results indicate that decision tree-based models outperformed the others. The Decision Tree technique was selected for the next step due to its superior performance compared to the Extra Trees Regressor, as evidenced by its lower MAE, MSE, and RMSE values. Moreover, its lower training time of 35x compared to the Extra Trees Regressor, as indicated by the TT(Sec) value, makes it a more efficient choice for the next stage of the training process.

Below are the hyperparameters selected in the CPU Model tuning step:

\begin{center}
\medskip
\fbox{\begin{minipage}{15em}
    \centering
    \textbf{DecisionTreeRegressor} (ccp\_alpha=0.0, criterion=`mse',\\ min\_impurity\_decrease=0.0,\\
    min\_impurity\_split=None, min\_samples\_leaf=1,\\
    min\_samples\_split=2,\\
    min\_weight\_fraction\_leaf=0.0,\\ random\_state=123,\\ splitter=`best')%
\end{minipage}}
\medskip
\end{center}

Table~\ref{Tab:dt-rtt} displays the result of the second training step for the selected model. One can see a performance improvement for most metrics.
\begin{table}[htb]
\centering
\begin{tabular}{|ccccccc|}
\hline
\multicolumn{7}{|c|}{\textbf{Decision Tree}} \\ \hline
\multicolumn{1}{|c|}{\textbf{fold}} & \multicolumn{1}{c|}{\textbf{MAE}} & \multicolumn{1}{c|}{\textbf{MSE}} & \multicolumn{1}{c|}{\textbf{RMSE}} & \multicolumn{1}{c|}{\textbf{R2}} & \multicolumn{1}{c|}{\textbf{RMSLE}} & \textbf{MAPE} \\ \hline
\multicolumn{1}{|c|}{\textbf{0}} & \multicolumn{1}{c|}{0.3036} & \multicolumn{1}{c|}{0.9071} & \multicolumn{1}{c|}{0.9524} & \multicolumn{1}{c|}{0.9992} & \multicolumn{1}{c|}{0.0846} & 0.0391 \\ \hline
\multicolumn{1}{|c|}{\textbf{1}} & \multicolumn{1}{c|}{0.3022} & \multicolumn{1}{c|}{0.8969} & \multicolumn{1}{c|}{0.9470} & \multicolumn{1}{c|}{0.9992} & \multicolumn{1}{c|}{0.0832} & 0.0393 \\ \hline
\multicolumn{1}{|c|}{\textbf{2}} & \multicolumn{1}{c|}{0.3058} & \multicolumn{1}{c|}{0.9244} & \multicolumn{1}{c|}{0.9614} & \multicolumn{1}{c|}{0.9992} & \multicolumn{1}{c|}{0.0847} & 0.0394 \\ \hline
\multicolumn{1}{|c|}{\textbf{3}} & \multicolumn{1}{c|}{0.3015} & \multicolumn{1}{c|}{0.8934} & \multicolumn{1}{c|}{0.9452} & \multicolumn{1}{c|}{0.9992} & \multicolumn{1}{c|}{0.0837} & 0.0389 \\ \hline
\multicolumn{1}{|c|}{\textbf{4}} & \multicolumn{1}{c|}{0.3030} & \multicolumn{1}{c|}{0.9103} & \multicolumn{1}{c|}{0.9541} & \multicolumn{1}{c|}{0.9992} & \multicolumn{1}{c|}{0.0836} & 0.0387 \\ \hline
\multicolumn{1}{|c|}{\textbf{Mean}} & \multicolumn{1}{c|}{\textbf{0.3032}} & \multicolumn{1}{c|}{\textbf{0.9064}} & \multicolumn{1}{c|}{\textbf{0.9520}} & \multicolumn{1}{c|}{\textbf{0.9992}} & \multicolumn{1}{c|}{\textbf{0.0840}} & \textbf{0.0391} \\ \hline
\multicolumn{1}{|c|}{\textbf{SD}} & \multicolumn{1}{c|}{0.0015} & \multicolumn{1}{c|}{0.0109} & \multicolumn{1}{c|}{0.0057} & \multicolumn{1}{c|}{0.0000} & \multicolumn{1}{c|}{0.0006} & 0.0003 \\ \hline
\end{tabular}
\caption{Decision Tree-based RTT Model}
\label{Tab:dt-rtt}
\end{table}

\paragraph{\textbf{Discussion}}
\medskip

These results show models that an administrator can use both models to estimate CPU or RTT usage in a cloud virtualization scenario. This type of insight assists in selecting the type of virtualization or driver technologies to help understand the trade-offs among the different configurations. Even when a scenario operates an RTT measurement frequency other than the three ones in the dataset, such as a frequency of 100Kpps, the model is able to accurately estimate CPU usage and RTT ping for such scenarios. This is important for decision-making without having to execute the experiments again. The methodology used in the data pre-processing allows these to be in a format suitable for machine learning so that the algorithms applied can generate a meaningful diagnosis. Available and post-processed data were used to create models based on various regression techniques. The best models used methods based on decision trees, mainly those that combine several decision trees to determine the result. This is primarily because these algorithms work well on structured data \cite{Chen:2016:XST:2939672.2939785}. These algorithms use multiple base learners, allowing them to learn complex relationships between resources present in the database used in training.

Note that other models can be developed based on the data provided in this work. Classification models can be applied; if necessary, a user can append to (augment) the database information from other scenarios and combinations, allowing further analysis.
\section{Conclusion and Future Works}
\label{sec:conc}
This work presents a study of the impact of virtualization on performance metrics, Round-Trip Time (RTT), and CPU Usage in a Cloud computing environment. A testbed was implemented, and experiments were conducted to obtain a dataset relating RTT and ``CPU Usage`` to the technological characteristics of the scenario, such as virtualization technology, driver type, CPU Affinity, frequency of RTT measurement packets, Load location (sender versus responder), and number of VMs running in the background.

The obtained data were subjected to pre-processing and various data science techniques, including statistical and data analyses, to draw conclusions and observe underlying patterns. The conclusions highlight that using Docker on the responder host decreases CPU usage. The rtl8139 driver worsens KVM virtualization performance compared to other drivers. In contrast, the VirtIO driver improves it, and a higher frequency in RTT measurement packets implies a lower end-to-end latency.

The dataset was used to create two regression models based on decision trees to estimate RTT and CPU Usage according to the characteristics and parameters of the network. These models help Cloud administrators decide when to use virtualization technology or drivers given a scenario and its configuration. An example is when the administrator needs to estimate CPU Usage and RTT for an I/O-intensive load application. By feeding the scenario parameters to the models, the administrator can determine if the results are compatible with the restrictions or if other configurations need to be tested.

The methodology used in creating the testbed and the experiments is available, allowing it to be replicated to obtain new data using other scenarios. Also, the dataset obtained and used in this work is publiclly available at \cite{6459234} and can be used by anyone who wishes to conduct new evaluations or develop other models.

Finally, other scenarios will be generated to obtain more data in future work using virtualization technologies not considered by this work, such as VMware, XEN, and OpenVZ. 
\bibliographystyle{IEEEtran}
\bibliography{bib} 

@inproceedings{Chen:2016:XST:2939672.2939785,
 author = {Chen, Tianqi and Guestrin, Carlos},
 title = {{XGBoost}: {A Scalable Tree Boosting System}},
 booktitle = {Proceedings of the 22nd ACM SIGKDD International Conference on Knowledge Discovery and Data Mining},
 series = {KDD '16},
 year = {2016},
 isbn = {978-1-4503-4232-2},
 location = {San Francisco, California, USA},
 pages = {785--794},
 numpages = {10},
 url = {http://doi.acm.org/10.1145/2939672.2939785},
 doi = {10.1145/2939672.2939785},
 acmid = {2939785},
 publisher = {ACM},
 address = {New York, NY, USA},
}

@article{BELLO2021103441,
title = {Cloud computing in construction industry: Use cases, benefits and challenges},
journal = {Automation in Construction},
volume = {122},
pages = {103441},
year = {2021},
issn = {0926-5805},
doi = {https://doi.org/10.1016/j.autcon.2020.103441},
url = {https://www.sciencedirect.com/science/article/pii/S0926580520310219},
author = {Sururah A. Bello and Lukumon O. Oyedele and Olugbenga O. Akinade and Muhammad Bilal and Juan Manuel {Davila Delgado} and Lukman A. Akanbi and Anuoluwapo O. Ajayi and Hakeem A. Owolabi}
}

@article{FILHO202273,
title = {An experimental investigation of Round-Trip Time and virtualization},
journal = {Computer Communications},
volume = {184},
pages = {73-85},
year = {2022},
issn = {0140-3664},
doi = {https://doi.org/10.1016/j.comcom.2021.12.006},
url = {https://www.sciencedirect.com/science/article/pii/S0140366421004722},
author = {Assis T. de Oliveira Filho and Eduardo Freitas and Pedro R. X. do Carmo and Djamel H.J. Sadok and Judith Kelner},
}

@InProceedings{10.1007/978-3-030-05057-3_5,
author="Du, Qingfeng
and He, Yu
and Xie, Tiandi
and Yin, Kanglin
and Qiu, Juan",
editor="Vaidya, Jaideep
and Li, Jin",
title="An Approach of Collecting Performance Anomaly Dataset for NFV Infrastructure",
booktitle="Algorithms and Architectures for Parallel Processing",
year="2018",
publisher="Springer International Publishing",
address="Cham",
pages="59--71",
isbn="978-3-030-05057-3"}

@InProceedings{7095802,  author={Felter, Wes and Ferreira, Alexandre and Rajamony, Ram and Rubio, Juan},  booktitle={2015 IEEE International Symposium on Performance Analysis of Systems and Software (ISPASS)},   title={An updated performance comparison of virtual machines and Linux containers},   year={2015},  volume={},  number={},  pages={171-172},  doi={10.1109/ISPASS.2015.7095802}}

@book{cohen2014applied,
  title={Applied multiple regression/correlation analysis for the behavioral sciences},
  author={Cohen, Patricia and West, Stephen G and Aiken, Leona S},
  year={2014},
  publisher={Psychology press}
}

@inproceedings{jain2020overview,
  title={Overview and importance of data quality for machine learning tasks},
  author={Jain, Abhinav and Patel, Hima and Nagalapatti, Lokesh and Gupta, Nitin and Mehta, Sameep and Guttula, Shanmukha and Mujumdar, Shashank and Afzal, Shazia and Sharma Mittal, Ruhi and Munigala, Vitobha},
  booktitle={Proceedings of the 26th ACM SIGKDD International Conference on Knowledge Discovery \& Data Mining},
  pages={3561--3562},
  year={2020}
}

@book{geron2019hands,
  title={Hands-on machine learning with Scikit-Learn, Keras, and TensorFlow: Concepts, tools, and techniques to build intelligent systems},
  author={G{\'e}ron, Aur{\'e}lien},
  year={2019},
  publisher={" O'Reilly Media, Inc."}
}

@article{zar2014spearman,
  title={Spearman rank correlation: overview},
  author={Zar, Jerrold H},
  journal={Wiley StatsRef: Statistics Reference Online},
  year={2014},
  publisher={Wiley Online Library}
}

@Manual {PyCaret, 
    title = {PyCaret: An open source, low-code machine learning library in Python}, 
    author = {Moez Ali}, 
    year = {2020}, 
    month = {April}, 
    note = {PyCaret version 1.0}, 
    url = {https://www.pycaret.org}
    }

@article{abdi2010principal,
  title={Principal component analysis},
  author={Abdi, Herv{\'e} and Williams, Lynne J},
  journal={Wiley interdisciplinary reviews: computational statistics},
  volume={2},
  number={4},
  pages={433--459},
  year={2010},
  publisher={Wiley Online Library}
}

@Article{electronics9081295,
AUTHOR = {Ahmed, Mohiuddin and Seraj, Raihan and Islam, Syed Mohammed Shamsul},
TITLE = {The k-means Algorithm: A Comprehensive Survey and Performance Evaluation},
JOURNAL = {Electronics},
VOLUME = {9},
YEAR = {2020},
NUMBER = {8},
ARTICLE-NUMBER = {1295},
URL = {https://www.mdpi.com/2079-9292/9/8/1295},
ISSN = {2079-9292}
}

@dataset{6459234,
  author       = {Pedro R. X. do Carmo and
                  Eduardo Freitas and
                  Assis T. de Oliveira Filho and
                  Djamel F H Sadok},
  title        = {A Round-Trip Time and Virtualization dataset},
  month        = 04,
  year         = 2022,
  publisher    = {Zenodo},
  version      = {v1-kvm-lxc-docker},
  doi          = {10.5281/zenodo.6459234},
  url          = {https://doi.org/10.5281/zenodo.6459234}
}
\end{document}